\documentclass[12pt,a4paper,final]{iopart}
\usepackage{cite}
\usepackage{subcaption}
\usepackage{graphicx}
\usepackage{color}

\expandafter\let\csname equation*\endcsname\relax

\expandafter\let\csname endequation*\endcsname\relax

\usepackage{amsmath}
\usepackage[breaklinks=true,colorlinks=true,linkcolor=blue,urlcolor=blue,citecolor=blue]{hyperref}

\newcommand{\EQ}{\begin{equation}}
\newcommand{\EN}{\end{equation}}
\newcommand{\be}{\begin{equation}}
\newcommand{\ee}{\end{equation}}
\newcommand{\bea}{\begin{eqnarray}}
\newcommand{\eea}{\end{eqnarray}}

\begin{document}

\title[Spectral densities of an active Brownian motion]{Spectral density of individual trajectories of an active Brownian particle}

\author{Alessio Squarcini$^{1,2}$, Alexandre Solon$^3$ \& Gleb Oshanin$^3$}
\address{$^1$Max-Planck-Institut f\"ur Intelligente Systeme,
Heisenbergstr. 3, D-70569, Stuttgart, Germany}
\address{$^2$ IV. Institut f\"ur Theoretische Physik, Universit\"at Stuttgart,
Pfaffenwaldring 57, D-70569 
Stuttgart, Germany}
\address{$^3$Sorbonne Universit\'e, CNRS, Laboratoire de Physique Th\'eorique de la Mati\`ere Condens\'ee (UMR CNRS 7600), 4 Place Jussieu, 75252 Paris Cedex 05, France}

\begin{abstract}
  We study analytically the single-trajectory spectral density (STSD)
  of an active Brownian motion as exhibited, for example, by the
  dynamics of a chemically-active Janus colloid. We evaluate the
  standardly-defined spectral density, {\it i.e.} the STSD averaged
  over a statistical ensemble of trajectories in the limit of an
  infinitely long observation time $T$, and also go beyond the
  standard analysis by considering the coefficient of variation
  $\gamma$ of the distribution of the STSD.  Moreover, we analyse the
  finite-$T$ behaviour of the STSD and $\gamma$, determine the
  cross-correlations between spatial components of the STSD, and
  address the effects of translational diffusion on the functional
  forms of spectral densities. The exact expressions that we obtain
  unveil many distinctive features of active Brownian motion compared
  to its passive counterpart, which allow to distinguish between these
  two classes based solely on the spectral content of individual
  trajectories.
 \end{abstract}


\maketitle



\section{Introduction}

Active matter encompasses a variety of systems that are driven locally
out of equilibrium, at the scale of each constituents. Most commonly,
these active particles consume energy to self-propel and are found
across scales, from macroscopic animals forming, {\it e.g.}, bird
flocks~\cite{ballerini} or fish schools~\cite{katz} to the microscopic
world where motile bacteria~\cite{berg} or self-propelled colloids
operate~\cite{paxton,howse}. The strong nonequilibrium driving is
responsible for a host of collective behaviours with no equilibrium
counterpart such as flocking~\cite{chate}, motility-induced
phase separation~\cite{tailleur} or the so-called bacterial
turbulence~\cite{dombrowski} to name just a few prominent ones.

Active systems have received considerable attention from physicists in
the last few decades for the theoretical and practical challenges that
they pose~\cite{clemens,solon2}. Even at the level of a single active
particle, one encounters a rich phenomenology with non-Boltzmann
stationary states~\cite{szamel,fodor,ginot}, complex interactions
with boundaries~\cite{lauga,popescu2} and the generic impossibility to
define state functions such as pressure~\cite{solon} or
temperature~\cite{solon3}.

Two main models have been used for the motion of an active
particle. In a first one, run-and-tumble particles (RTPs) alternate
periods of directional swimming with short-duration ``tumbles'' during
which they randomise their direction of motion. That model accounts
for the trajectories of bacteria like {\it Escherichia coli}
\cite{berg,schnitzer}. In the other class of models, particles
self-propel in a direction that fluctuates because of rotational
noise. These active Brownian particles (ABPs)~\cite{fily} are
appropriate to describe, for example, Janus colloidal particles
decorated with a catalytic patch which prompts a chemical reaction in
the surrounding solution~\cite{palacci,buttinoni}. The reaction then
generates gradients of product species across the surface of the
particle leading to a non-zero self-propulsive force $F$ (see,
e.g. Refs.~\cite{ramin1,osh}). Within the last decade, various
characteristics including the position probability density functions,
the marginal distributions and the first-passage properties have been
studied for both
RTPs~\cite{angel,detch,malakar,beni,dhar,santra,mori,basu3,sin,raich}
and ABPs~\cite{kurzthaler,pototsky,sevilla,basu1,baruch,santra2}.  In
addition, note that many other models of active particles have been
proposed~\cite{ebeling,martin,pavol}.

Here, we will focus on the dynamics of an active Brownian particle in
the two-dimensional $(x,y)$-plane (see Fig. \ref{fig_PSDfinite}),
which is realised experimentally, e.g., in a situation in which a
Janus colloid has sedimented on a bottom plate or is trapped at the
interface separating two distinct liquids~\cite{col1,col2} and is
therefore restricted to 2d motion.  For simplicity, we assume
that the rotational diffusion also takes place in the plane which
may or may not be the case depending on the experimental
setup. Finally, we neglect translational (passive) diffusion which is
often very small compared to the active motion. Within these
approximations, the system of equations describing the time-evolution
of the components $x(t)$ and $y(t)$ of the position of an ABP can be
written\footnote{We note in passing that exactly the same
  mathematical model emerges within the context of edge-detection in
  computer vision \cite{m}. We thank U. Basu for bringing this work to
  our attention.}
\begin{equation}
\begin{aligned}
\label{02}
\dot{x}(t) & = v \cos \theta(t) \, ,  \quad x(0) = 0 \,, \\
\dot{y}(t) & = v  \sin \theta(t) \, , \quad y(0) =0 \,, 
\end{aligned}
\end{equation}
where the dot denotes time derivative, $v$ is the constant
self-propulsion velocity and $\theta(t)$ is the instantaneous polar
angle defining the direction of motion, {\it i.e.} the angle between
the velocity vector and the $x$-axis.  For an ABP, $\theta$
changes due to rotational diffusion so that this process is a
one-dimensional Brownian motion (BM) with zero mean and covariance
\begin{equation}
\label{01a}
\langle \theta(t_1) \theta(t_2) \rangle = \frac{2}{\tau_R}  \, {\rm min}\left(t_1, t_2\right) \,,
\end{equation}
with $\tau_R$ the persistence time of the particle, the characteristic
time after which the initial orientation is forgotten.  Here and in
the rest of the paper, the angle brackets denote averaging with
respect to different realisations of $\theta(t)$ and we choose the
initial condition $\theta(t=0)=0$.

Note that two salient features of the model in
Eqs.\eqref{02}-\eqref{01a} are that a) the effective noise is a
non-linear function of the random process $\theta(t)$ and b) the
components of the instantaneous velocity ($v_x(t) = \dot{x}(t)$ and
$v_y(t) = \dot{y}(t)$) obey
\begin{align}
\label{v}
v^2_{x}(t) + v^2_{y}(t) \equiv v^2
\end{align}
at any moment in time $t$, which shows explicitly that the components
are coupled. These two circumstances clearly entail departures from a
standard BM behaviour at some transient stages (see, e.g.,
Refs. \cite{pototsky,sevilla,bar,basu1,baruch,santra2} and
below). However, as can be expected intuitively, on large time scales
the ABP follows a standard two-dimensional BM with an effective
diffusion coefficient $D_e=v^2\tau_R/2$~\cite{catesEPL}. Note that it remains true even
if the particle is placed in an external potential, provided the
variations of the potential are small on the scale of the persistence
length $l_p=v\tau_R$~\cite{solon3}.

\begin{figure}[htbp]
\centering
\includegraphics[width=90mm]{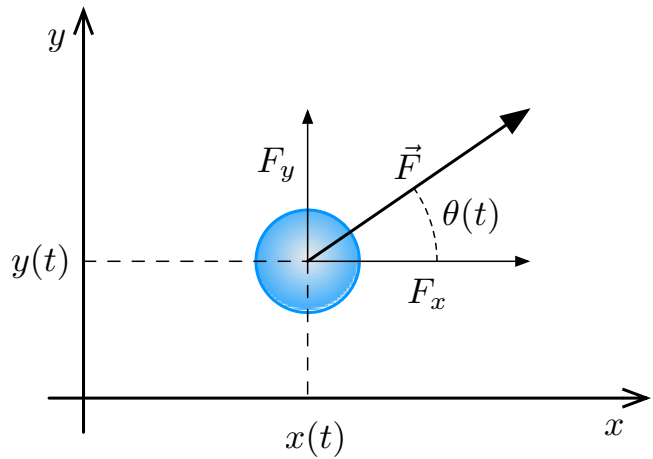}
\caption{An active colloid on a two-dimensional $(x,y)$-plane. The
  vector $\vec{F}$ (with components $F_x$ and $F_y$) denotes the
  self-propulsion force emerging due to interactions with the
  environment. $x(t)$ and $y(t)$ are the coordinates of the
  instantaneous position of the centre of the colloid.}
\label{fig_PSDfinite}
\end{figure}

In this paper, we are interested in the power spectral density of an
ABP, an aspect that has heretofore not been investigated. It is
well-known that the power spectral density of any stochastic process
embodies a wealth of information on its temporal evolution and
correlations. As such, it is one of the most widely used
characterisation tools, and nowadays a substantial knowledge about
spectral densities of a variety of processes has been accumulated (see,
e.g., Refs. \cite{flandrin,eli,eli2,dechant,olivier,dean,flyv,satya,alessio,eli3}
and references therein).

According to the standard definition, the power spectral density
$\mu_x(f)$ of the process $x(t)$ at frequency $f$ involves taking a
statistical average over realisations of the process and the limit of
an infinitely long trajectory. For a single trajectory of duration
$T$, one defines the single trajectory spectral density (STSD)
\begin{equation}
\label{def2}
S_x(f,T) = \frac{1}{T} \left| \int_0^T \textrm{d}t \, e^{i f t} \, x(t) \right|^2 \,,
\end{equation} 
which is a functional of the trajectory and itself a random
process. We denote its average by
$\mu_x(f, T) = \langle S_x(f,T) \rangle$ so that the spectral density
is obtained as 
\begin{equation}
\label{defs1}
\mu_x(f) = \lim_{T \to \infty} \mu_x(f, T) \,.
\end{equation}

We note that a standard textbook analysis focuses precisely on
$\mu_x(f)$ which is an ensemble-averaged property that can be
understood as a suitable Fourier transform of the covariance function
of the process $x(t)$.  In some instances, however, the limit
$T \to \infty$ does not exist (see, {\it e.g.},
Refs. \cite{flandrin,eli,eli2,dechant,olivier,dean,satya,alessio,eli3}
for some examples) and one has to resort to either alternative
definitions of the spectral density ({\it e.g.} the one due to Wigner
and Ville, see Refs. \cite{flandrin,alessio} and references therein),
or operate directly with the finite-$T$ counterpart $\mu_x(f,
T)$. Indeed, this quantity is well-defined for any finite $T$ and also
gives access to useful information about the ageing properties
($T$-dependence) of spectral densities
\cite{krapf1,krapf2,sposini1,sposini2,carlos,sara}.  In particular,
$\mu_x(f, T)$ evaluated at zero frequency reads
\begin{equation}
\label{area}
\mu_x(f = 0, T) = \frac{1}{T} \left\langle \left(\int^T_0 \textrm{d}t \, x(t)\right)^2 \right\rangle \,,
\end{equation}
which is the averaged squared area under random curve $x(t)$ on the
interval $t \in (0,T)$ divided by $T$. Observing the rate at which
$\mu_x(f = 0, T) $ diverges, when it does, as $T$ tends to infinity,
tells us much about the process
itself~\cite{satya,krapf1,krapf2,sposini1,sposini2}.

Going beyond the standard definition, recent works investigated the
statistical properties of the STSD $S_x(f,T)$ and calculated its full
probability density function for several Gaussian
processes~\cite{krapf1,krapf2,sposini1,sposini2}. It was realised that
the coefficient of variation of this distribution, defined formally by
\begin{equation}
\label{gamma}
\gamma = \gamma(f,T) = \frac{\sqrt{\langle S^2_x(f,T)\rangle - \mu^2_x(f,T)}}{\mu_x(f,T)} \,,
\end{equation}
is an important characteristic parameter. By definition, it measures
the relative amplitude of fluctuations of the STSD at given $f$ and
$T$ with respect to its mean value, and hence, shows how
representative of the actual behaviour $\mu_x(f, T) $ is. It was shown
in Refs.~\cite{krapf1,krapf2,sposini1,sposini2} that for several known
Gaussian processes the coefficient of variation obeys for any $f$ and
$T$ a two-sided inequality: $1 \leq \gamma \leq \sqrt{2}$, implying
that fluctuations of the STSD are generically bigger than the mean
value, and hence that large statistical samples are necessary to
reliably evaluate $\mu_x(f, T) $ in practice. Next, even when the
asymptotic behaviours of $\mu_x(f)$ are the same for distinctly
different processes, it appears that the values of $\gamma$ may be
very different, thus allowing to distinguish between different
processes. In particular, for the so-called fractional BM (fBM) with
Hurst index $H$ --- a family of anomalous diffusions --- $\mu_x(f) \sim 1/f^2$ in the limit $f \to \infty$ for standard BM
  $(H = 1/2)$ as well as any super-diffusive fBm with
  $H > 1/2$~\cite{krapf2}. On the contrary, $\gamma$ behaves
  differently in the two cases: as $f \to \infty$, it approaches a
  universal value $\sqrt{2}$ for super-diffusion and the value
  $\sqrt{5}/2$ for standard BM. For sub-diffusive fractional BM the
  behaviour is again different: $\gamma \to 1$ as
  $f \to \infty$~\cite{krapf2}. Correspondingly, the analysis of
$\gamma$ (in addition to the more commonly used analysis of the
mean-squared displacement), provides a robust criterion for anomalous
diffusion.

At present, neither the form of the standard power spectral density
$\mu_x(f)$, nor its ageing properties or the behaviour of the
characteristic parameter $\gamma$ are known for the much popular ABP
model.  We bridge this gap by evaluating explicit expressions
for these characteristic properties, discuss their similarity to those
of standard BM and emphasise several distinctive features.  We also
address additional questions about the cross-correlations of the STSD
for the $x$- and $y$-components, and about the effects of
translational diffusion.  The paper is organised as follows: in
Sec.~\ref{1} we first evaluate the exact form of the standard textbook
power spectral density of trajectories of an ABP in the limit of an
infinite observation time before discussing some features of the
finite-$T$ case. Sec.~\ref{2} is devoted to the analysis of the coefficient
of variation $\gamma$ associated with the random functionals
$S_x(f,T)$ and $S_y(f,T)$ both in the limit $T \to \infty$ and for
finite $T$. We also analyse at the end of Sec.~\ref{2} the behaviour
of the Pearson correlation coefficient for $S_x(f,T)$ and
$S_y(f,T)$. In Sec.~\ref{3} we study the effects of translational
diffusion on the STSD before concluding in Sec.~\ref{4} with a brief
recapitulation of our results.

\section{Power spectral density of trajectories of an active Brownian particle}
\label{1}

\subsection{Position correlation functions}
Computing the standard power spectral density $\mu_x(f)$ and
$\mu_y(f)$ necessitates only the knowledge of the two-time correlation
functions $C_{xx}(t_1,t_2)=\langle x(t_{1}) x(t_{2}) \rangle$ and
$C_{yy}(t_1,t_2)=\langle y(t_{1}) y(t_{2}) \rangle$. These can be
determined directly from Eqs.\eqref{02}-\eqref{01a}, which we show
explicitly here for $x(t)$.

We first differentiate $C_{xx}$ about $t_1$ and $t_2$ giving
\begin{equation}
  \label{eq:Cxxderive}
  \frac{\partial^2C_{xx}(t_1,t_2)}{\partial t_1\partial t_2}=v^2\langle\cos\theta(t_1)\cos\theta(t_2)\rangle \, .
\end{equation}
Applying the It\=o formula~\cite{gardiner} to differentiate $\cos^2\theta(t)$ we obtain
\begin{equation}
  \label{eq:costheta}
  \frac{d}{dt}\langle \cos^2\theta(t)\rangle=-\frac{\langle 2-4\cos^2\theta(t)\rangle}{\tau_R}
\end{equation}
which integrates to
\begin{equation}
  \label{eq:costheta2}
  \langle \cos^2\theta(t)\rangle=\frac{1}{2}\left(1+e^{-4t/\tau_R}\right) \, .
\end{equation}
Applying again the It\=o formula to differentiate $\cos\theta(t_1)\cos\theta(t_2)$ with respect to $t_1$ and integrating
gives for the correlator
\begin{equation}
  \label{eq:cost1t2}
  \langle\cos\theta(t_1)\cos\theta(t_2)\rangle=\frac{1}{2}\left(1+e^{-4t/\tau_R}\right)e^{-(t_1-t_2)/\tau_R}.
\end{equation}
Finally, integrating Eq.~(\ref{eq:cost1t2}) over $t_1$ and $t_2$ gives
the correlation for $x$ (and similarly for $y$)
\begin{equation}
\begin{aligned}
\label{08}
C_{xx}(t_1,t_2)  = v^{2} \tau_{R}^{2} \biggl[\frac{t_{2}}{\tau_R} - \frac{1}{4} + \frac{1}{3} e^{-t_{1}/\tau_R} + \frac{1}{3} e^{-t_{2}/\tau_R} - \frac{1}{12} e^{-4t_{2}/\tau_R} \\- \frac{1}{2} e^{-(t_{1}-t_{2})/\tau_R}
+ \frac{1}{6} e^{-(t_{1}+3t_{2})/\tau_R} \biggr] \,,  \\
C_{yy}(t_1,t_2)   = v^{2} \tau_{R}^{2} \biggl[ \frac{t_{2}}{\tau_R} - \frac{3}{4} + \frac{2}{3} e^{-t_{1}/\tau_R} + \frac{2}{3} e^{-t_{2}/\tau_R} + \frac{1}{12} e^{-4 t_{2}/\tau_R} \\
 - \frac{1}{2} e^{-(t_{1}-t_{2})/\tau_R} - \frac{1}{6} e^{-(t_{1}+3t_{2})/\tau_R} \biggr]  \,,
\end{aligned}
\end{equation}
where we assumed without loss of generality that $t_1 \geq t_2$. One
can also directly read off from Eqs.\eqref{08} the exact expressions for
the mean-squared displacements of the components at time $t$:
\begin{equation}
\begin{aligned}
\label{09}
\langle x^{2}(t) \rangle &= v^2 \tau_R^2 \biggl[ \tau - \frac{3}{4} + \frac{2}{3} e^{-\tau} + \frac{1}{12} e^{-4 \tau} \biggr]  \,, \\
\langle y^{2}(t) \rangle &=  v^2 \tau_R^2 \biggl[ \tau - \frac{5}{4} + \frac{4}{3} e^{-\tau} - \frac{1}{12} e^{-4 \tau} \biggr] \, ,
\end{aligned}
\end{equation}
where we introduced the dimensionless time variable $\tau = t/\tau_R$.

From Eq.~(\ref{09}), we see that, at long times $t/\tau_R \to \infty$,
the differences between $x$ and $y$ due to the initial condition are
forgotten and both $\langle x^{2}(t) \rangle$ and
$\langle y^{2}(t) \rangle$ grow linearly in time, indicating a
diffusive motion with an effective diffusion coefficient 
\begin{align}
\label{diff}
D_e = \frac{v^2 \tau_R}{2} \,,
\end{align}
as derived previously in \cite{catesEPL}.
At short times, $x(t)$ and $y(t)$ behave differently:
$\langle x^{2}(t) \rangle \simeq v^2 t^2$, {\it i.e.} it exhibits a simple
ballistic motion with the velocity $v$, while
$\langle y^{2}(t) \rangle$ grows faster with time at this initial
stage, $\langle y^{2}(t) \rangle \simeq 2 v^2 t^3/(3 \tau_R)$.

Plugging the expressions of the correlators Eq.\eqref{08} into
Eq.~\eqref{def2} and performing the two-fold integral allows to
compute $\mu_x(f, T)$ and $\mu_y(f, T)$ explicitly. We give this exact
form in the infinite-trajectory limit in the next subsection. We omit
the lengthy expression for arbitrary $T$ but discuss limiting
situations at finite $T$ in Sec.~\ref{finite}.

\subsection{Infinite-$T$ limit.}
\label{infinity}

In the limit of an infinitely long observation time, we obtain the
following exact expression for the power spectral density defined in
Eq.~(\ref{defs1})
\begin{align}
\label{tb}
\mu_x(f) = \mu_y(f) = \frac{2 D_e}{f^2} + \frac{2 D_e}{f^2}  \frac{1}{1 + \tau_R^2 f^2} \,, \quad \left[\mu^{(BM)}(f) = \frac{4 D}{f^2}\right]  \,, 
\end{align}  
where inside the brackets we give for comparison the same quantity for
a standard BM with diffusion coefficient $D$.  Note that the
difference in the evolution of $x(t)$ and $y(t)$ at short times does
not play any role in the limit $T \to \infty$, so that,
  in this limit, the power spectral densities are equal for the two
  components for any $f$.

Unsurprisingly, the form of the power spectral density presented in
Eq.~(\ref{tb}) is more complicated for an ABP than for standard BM and
consists of two terms. The first one $2 D_e/f^2$ is such that it would
be generated by BM with a diffusion coefficient $D_e/2$. The second
one is a product of the standard Brownian result and a Lorentzian
function, the latter being a generic feature of dynamics in presence
of a constant restoring force, as observed with the Ornstein-Uhlenbeck
process (see e.g. Ref.~\cite{flyv}) or for a BM with stochastic
reset~\cite{satya}.
Our analysis adds to this list and shows that the Lorentzian form is
rather universal for the power spectral density of active processes.
In the asymptotic limit $f \to 0$ (which corresponds to the large-$t$
asymptotic behaviour of $x(t)$), the two terms equally contribute
$2 D_e/f^2$, to give the result expected for BM with a diffusion
coefficient $D_e$. In the opposite limit $f \to \infty$, the second
term in Eq.\eqref{tb} vanishes at a faster rate ($\sim 1/f^4$) than
the first one ($\sim 1/f^2$) which then dominates. The power spectral
density in that limit is thus the same as for BM with a diffusion
coefficient $D_e/2$.

\subsection{$T$-dependent behaviour of  $\mu_x(f,T)$ and $\mu_y(f,T)$.}
\label{finite}

Let us first consider the case $f = 0$. It is of special interest
because, in this limit, the STSD becomes the squared area under the
random curve drawn by a realisation $x(t)$. Moments of areas under
random curves, as well the moments of areas conditioned on some
extreme events have recently received a lot of attention (see, {\it
  e.g.}  Refs.~\cite{satya,krapf2,sposini1,sat,bar1,bar2,bar3} and
references therein) but, to the best of our knowledge, these
quantities have not been computed for active BM. In our case, while
the areas themselves evidently average to zero, the variances of the
areas, {\it i.e.} $\mu_x(f=0,T)$ and $\mu_y(f=0,T)$, are clearly
positive-definite increasing functions of $T$. Note that the case
$f = 0$ is somewhat peculiar: since the spectral densities
$\mu_x(f,\infty)$ and $\mu_y(f,\infty)$ diverge in this limit and thus
are not defined.  In fact, it is well-known that in the analysis of
spectral properties of random processes the limits $f \to 0$ and
$T \to \infty$ often cannot be interchanged, which renders the
situation somewhat subtle \cite{eli2,dechant} (see also below).

We present here the exact result for the $x$-component and an
arbitrary duration $T$, which reads
\begin{align}
\label{area}
\mu_x(f=0,T) &= \frac{1}{T} \left \langle \left(\int^T_0 \textrm{d}t \, x(t) \right)^2 \right \rangle =  \nonumber\\ &=v^2 \tau_R^3 \biggl[\frac{{\cal T}^2}{3}   - \frac{ {\cal T}}{4} - \frac{3}{8} + \frac{35}{32 {\cal T}} 
- \frac{2}{3} e^{-{\cal T}} - \frac{10}{9 {\cal T}} \, e^{-{\cal T}}+ \frac{5}{288 {\cal T}} \, e^{-4 {\cal T}}\biggr]\,, 
\end{align}
where ${\cal T} = T/\tau_R$ is the dimensionless observation
time. The corresponding expression for $\mu_y(f=0,T)$ has a similar
${\cal T}$-dependence and differs only by the values of numerical
factors.  From Eq.~\eqref{area} we find that, in the limit
$T \to \infty$, at leading order,
\begin{align}
\label{lead}
\left \langle \left(\int^T_0 \textrm{d}t \, x(t) \right)^2 \right \rangle = \frac{2 D_e}{3}T^3 + O\left(T^2\right) \,,
\end{align} 
where the symbol $O\left(T^2\right)$ indicates that the omitted
subdominant terms are of order $T^2$. The leading term in
Eq.~\eqref{lead} coincides exactly with the behaviour of the averaged
squared area under a standard BM with diffusion coefficient $D_e$. In
contrast, the short-$T$ behaviour is different,
\begin{align}
\left \langle \left(\int^T_0 \textrm{d}t \, x(t) \right)^2 \right \rangle = \frac{v^2}{4}T^4 + O\left(T^5\right) \,,
\end{align}
and is associated with the transient ballistic regime.
\begin{figure}[htbp]
\centering
\includegraphics[width=100mm]{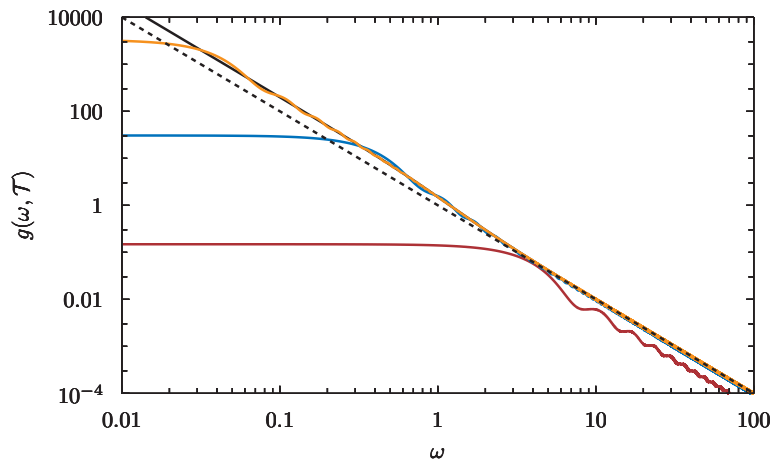}
\caption{Scaled power spectrum
  $g(\omega,{\cal T}) = \mu_x(f,T)/(v^2 \tau_R^3)$, Eq. \eqref{g}, as
  function of the dimensionless frequency $\omega = \tau_R f$ for
  three values of the scaled observation time ${\cal T} = T/\tau_R$:
  ${\cal T}=1$ (red), ${\cal T}=10$ (blue) and ${\cal T}=100$
  (orange). Solid (black) curve depicts the result in Eq.\eqref{tb}
  (${\cal T}=\infty$). As a guide for an eye, the dashed line is the
  function $1/\omega^{2}$.}
\label{fig_PSDfinite}
\end{figure}

Further on, let us assume that $f$ is strictly positive and analyse
how the infinite-$T$ form of Eq.~\eqref{tb} is approached when the
observation time $T$ increases. To this end, we first note that
$\mu_x(f,T)$ can be represented as
\begin{align}
\label{g}
\mu_x(f,T) = v^2 \tau_R^3 \, g(\omega, {\cal T}) \,,
\end{align}
where $\omega = \tau_R f$ is a dimensionless frequency and
${\cal T} = T/\tau_R$, as above, is a dimensionless observation
time. $g(\omega, {\cal T})$ is then a scaled power spectral
density. It can be computed explicitly but we omit here its lengthy
expressions.  In Fig.~\ref{fig_PSDfinite} we plot
$g(\omega, {\cal T})$ for three values of ${\cal T}$, together with
the limiting form in Eq. \eqref{tb}.  We observe that, as a function
of $\omega$, $g(\omega, {\cal T})$ generically shows a plateau at
sufficiently small values of $\omega$, followed rather abruptly by a
power-law decay $g(\omega, {\cal T})\sim 1/\omega^2$ at larger
$\omega$.  As ${\cal T}$ increases, the plateau appears at smaller
values of $\omega$ and is shifted towards larger values. Even for a
rather small value ${\cal T} = 1$, the power-law decay of
$g(\omega, {\cal T})$ appears very close to the ultimate
infinite-${\cal T}$ form \eqref{tb} (black solid line in
Fig.~\ref{fig_PSDfinite}), with a somewhat smaller amplitude.  For
larger observation times, (${\cal T} = 10$ and ${\cal T} = 100$), the
agreement is nearly perfect and starts from smaller values of the
scaled frequency. Unsurprisingly, the smaller $f$ is, the larger the
observation time needs to be to approach the asymptotic result.

Lastly, we evaluate the total power as a function of the observation
time. After straightforward calculations, we find
\begin{equation}
\begin{aligned}
\label{16}
\int_{0}^{\infty}\textrm{d}f \, \mu_{x}(f,T) & = \pi \frac{v^{2} \tau_R T}{2} \biggl[1 - \frac{3}{2 {\cal T}}  
+ \frac{33}{24 {\cal T}^2} -
\frac{4}{3 {\cal T}^2} e^{-{\cal T}} - \frac{1}{24 {\cal T}^2} e^{-4 {\cal T}} \biggr] \, , \\
\int_{0}^{\infty}\textrm{d}f \, \mu_{y}(f,T) & =  \pi \frac{v^{2} \tau_R T}{2} \biggl[1 - \frac{5}{2 {\cal T}}  
+ \frac{63}{24 {\cal T}^2}  -
\frac{8}{3 {\cal T}^2} e^{-{\cal T}} + \frac{1}{24 {\cal T}^2} e^{-4 {\cal T}} \biggr] \, ,
\end{aligned}
\end{equation}
which shows that the total power diverges linearly with $T$ in the limit $T \to \infty$.
  We note that, in general, in order for the functional $S_x(f,T)$ in Eq.\eqref{def2} to be meaningfully interpreted 
  as the power spectral density of a non-stationary process $x(t)$, it has to verify some conditions. In particular, in the asymptotic limit $T \to \infty$, 
  the total power as defined in the first line in Eqs.\eqref{16} must be equal, up to a factor $\pi$, to $\langle x^{2}(T) \rangle$~\cite{dechant}. We  observe that such a  condition indeed holds.

\section{Coefficient of variation and cross correlations}
\label{2}

We now go beyond the standard spectral analysis to focus on two
quantities that measure the level of fluctuations between realisations
of the STSD $S_x(f,T)$ and $S_y(f,T)$. The first one is the
coefficient of variation defined in Eq.\eqref{gamma} that quantifies
the relative amplitude of fluctuations in either $S_x(f,T)$ or
$S_y(f,T)$. The second is the Pearson coefficient $\rho_S$ that quantifies how correlated the values of
$S_x(f,T)$ and $S_y(f,T)$ are. It is defined as 
\begin{align}
\label{pearson}
\rho_{S} = \dfrac{\left \langle S_x(f,T) S_y(f,T) \right \rangle- \mu_x(f,T) \mu_y(f,T)}{\sqrt{\left[\left \langle S^2_x(f,T) \right \rangle - \mu^2_x(f,T)\right]\left[ \left \langle S^2_y(f,T) \right \rangle - \mu^2_y(f,T)\right]}} \,.
\end{align}
To this end, we need to calculate the second moments of $S_x(f,T)$ and
$S_y(f,T)$, and the cross moment
$\left \langle S_x(f,T) S_y(f,T) \right \rangle$.  Once again, this
can be done explicitly. The calculation itself is tedious but rather
straightforward so that we skip the details and present the final
results only.

\subsection{Infinite-$T$ limit.} 

\begin{figure}[htbp]
\centering
\includegraphics[width=100mm]{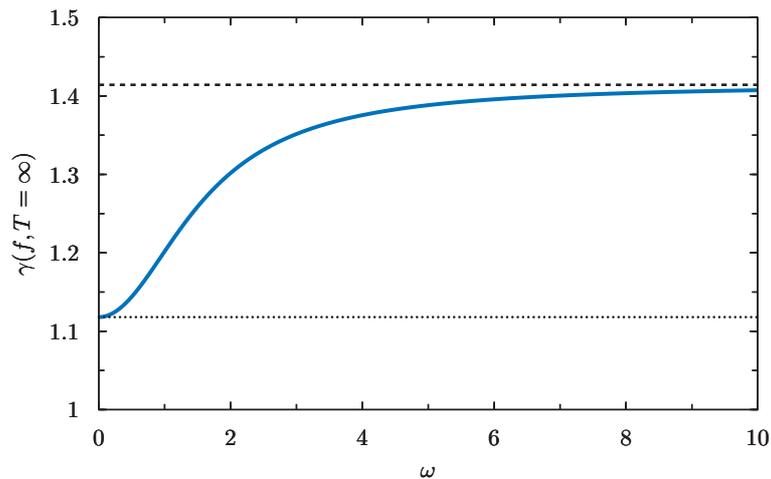}
\caption{The coefficient of variation $\gamma$ for infinite
  observation time Eq.\eqref{gamma_infinity} as function of
  $\omega = \tau_R f$. The dashed line indicates the asymptotic value
  $\sqrt{2}$, achieved in the limit $f \to \infty$, while the dotted
  line corresponds to $\sqrt{5}/2$, the value of the coefficient of
  variation for a standard BM at any frequency.}
\label{fig_gamma}
\end{figure}

In the limit of an infinitely large observation time, the second
moments obey
\begin{align}
\label{second}
\left \langle S^2_x(f,\infty)\right \rangle = \left \langle S^2_y(f,\infty)\right \rangle = \tau_R^2 v^4  \, \frac{\left(9 + 10 \tau_R^2 f^2 + 3 \tau_R^4 f^4\right)}{f^4 (1 + \tau_R^2 f^2)^2} \,,
\end{align}
which yields the following compact result for the coefficient of variation :
\begin{align}
\label{gamma_infinity}
\gamma(f,T=\infty) = \frac{\sqrt{5 + 6 \tau_R^2 f^2 + 2 \tau_R^4 f^4}}{2 + \tau_R^2 f^2} \,, \quad \left[\gamma^{(BM)}(f,T = \infty) = \frac{\sqrt{5}}{2}  \,\,\, \text{for any $f > 0$} \right] \,,
\end{align}
where the expression in the brackets is the same quantity computed for
a standard BM (see, Ref. \cite{krapf1} for more details). We depict
$\gamma$ as expressed in Eq.\eqref{gamma_infinity} in
Fig. \ref{fig_gamma} as a function of the dimensionless frequency
$\omega = \tau_R f$.

A few remarks are in order:
\begin{enumerate}

\item While $\gamma$ for standard BM is a constant independent of $f$
  ($ = \sqrt{5}/2$ for any $f>0$), for an ABP it is a monotonically
  increasing function of the frequency, when $T$ is infinitely large. It interpolates between
  $\sqrt{5}/2$, reached as expected in the limit $f \to 0$, and
  $\sqrt{2}$ achieved in the asymptotic limit $f \to \infty$. Hence,
  for any $f > 0$ we have
  $\gamma(f,T=\infty) > \gamma^{(BM)}(f,T = \infty)$, which highlights
  the difference between the two random processes.

\item $S_x(f,\infty)$ and $S_y(f,\infty)$ are strongly
  fluctuating. Indeed, since $\gamma(f,\infty) > 1$, the standard
  deviation of, say $S_x(f,\infty)$, is bigger than its mean value,
  $\mu_x(f,\infty)$, which implies that the latter is not
  representative of the actual behaviour.

\item The maximal value of $\gamma(f,\infty)$ is the same
  ($=\sqrt{2}$) as the one achieved in the limit $f \to \infty$ by the
  coefficient of variation of super-diffusive Gaussian processes ({\it
    e.g.} super-diffusive fractional BM \cite{krapf2} or
  super-diffusive scaled BM \cite{sposini1}). Such a coincidence is
  somewhat surprising, given the non-Gaussian nature of an ABP.
\end{enumerate}

\subsection{Finite-$T$ behaviour of the coefficient of variation}

\begin{figure}[htbp]
\centering
\includegraphics[width=100mm]{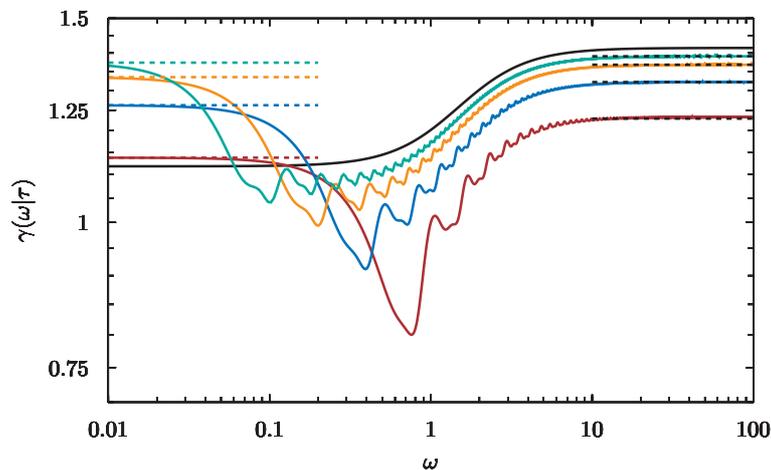}
\caption{The coefficient of variation $\gamma(f,T)$ for finite $T$ as
  function of $\omega = \tau_R f$ for four values of the reduced
  observation time: ${\cal T}=10$ (red), ${\cal T} =20$
  (blue), ${\cal T}=40$ (orange), and ${\cal T}=80$ (green).  The
  solid (black) curve is the asymptotic form
  Eq.~\eqref{gamma_infinity}, which corresponds to ${\cal T}=\infty$.
  The horizontal dashed lines are the asymptotic results in
  Eqs.~\eqref{asymp1} and \eqref{asymp2}.}
\label{fig_gammaT}
\end{figure}

Let us now consider the case of a finite observation time
$T$, a more realistic situation if one is to compare with experiments
or numerical simulations.  In Fig.~\ref{fig_gammaT} we depict
$\gamma(f,T)$ as function of the dimensionless frequency
$\omega = \tau_R f$ for several values of the dimensionless
observation time ${\cal T} = T/\tau_R$, together with the
infinite-${\cal T}$ limit of Eq.~\eqref{gamma_infinity} (solid black
curve).

We observe, interestingly enough, the non-commutation of limits
$T \to \infty$ and $f \to 0$ that we have mentioned previously so that
one obtains a completely different behaviour for $\gamma$ depending on
which of the limits is taken first. More precisely, the small-$f$
behaviour of $\gamma$ at finite $T$ does not approach
$\gamma(f,T=\infty)$ as $T$ is increased. The latter approaches a
constant value $\sqrt{5}/2$ as $f \to 0$ while the former tends to
some $T$-dependent value which drifts away from $\sqrt{5}/2$ as $T$
increases. This is again very different from the behaviour of
$\gamma^{(BM)}(f=0,T)$ for standard BM which also shows the
non-commutativity of limits but equals $\sqrt{2}$ {\it for any $T$} when
$f=0$~\cite{krapf1}.

The $T$-dependent expression $\gamma(f = 0,T)$ is rather
cumbersome. However, in the limit $T \to \infty$ it reduces to the
following asymptotic expansion
 \begin{align}
\label{asymp1}
\gamma(f=0,T) & = \sqrt{2} - \frac{189}{40 \sqrt{2} \, {\cal T}} + \frac{57369}{6400 \sqrt{2} \, {\cal T}^{2}} + O\left(\frac{1}{{\cal T}^{3}}\right) \, . 
\end{align}
Therefore, $\gamma(f = 0,T)$ attains the limiting value $\sqrt{2}$ of
standard BM but only in the limit $T \to \infty$ (whereas
$\gamma^{(BM)}(f = 0,T)=\sqrt{2}$ for any $T$). In the opposite limit
$f \to \infty$, $\gamma(f,T)$ approaches a limiting $T$-dependent
value, which is different from $\sqrt{2}$ attained at
$T\to\infty$. For large $T$, the asymptote becomes
\begin{align}
\label{asymp2}
\gamma(f =\infty,T) & = \sqrt{2} - \frac{21}{8 \sqrt{2} \, {\cal T}}  + \frac{55}{768 \sqrt{2} \, {\cal T}^2}  
+ O\left(\frac{1}{{\cal T}^{3}}\right) \, . 
\end{align}
One thus finds that, this time, the limits commute and one does
recover in the limit $T \to \infty$ the coefficient of variation
$\gamma(f = \infty,T = \infty)$ computed in
Eq.~\eqref{gamma_infinity}. The limits $f \to \infty$ and
$T \to \infty$ can thus be taken in either order.

\begin{figure}[htbp]
\centering
\includegraphics[width=100mm]{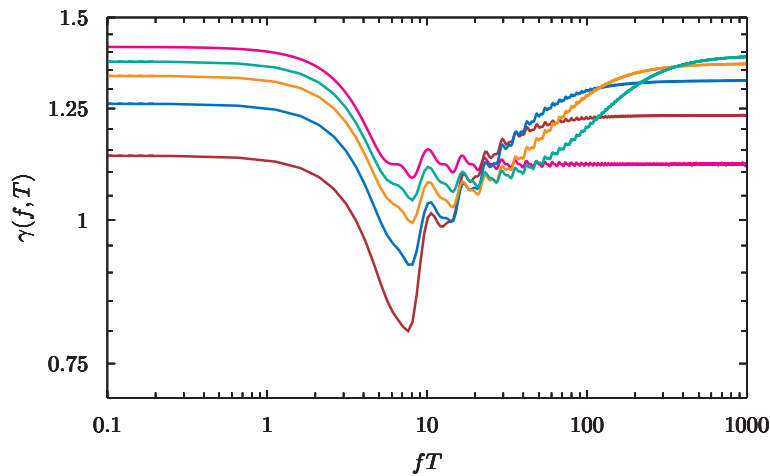}
\caption{The coefficients of variation $\gamma(f,T)$ and
  $\gamma^{(BM)}(f,T)$ (magenta) as functions of $f T$. The results
  for an ABP are presented for ${\cal T} =T/\tau_R =80$
  (green), ${\cal T}=40$ (orange), ${\cal T}=20$ (blue) and
  ${\cal T}=10$ (red).}
\label{fig_gammacfr}
\end{figure}

Further on, we present in Fig.~\ref{fig_gammacfr} a more detailed
comparison between the coefficients of variation for an ABP and a
standard BM. We recall that the latter is function of the product
$f T$ only~\cite{krapf1} and hence, we plot $\gamma(f,T)$ and
$\gamma^{(BM)}(f,T)$ as functions of $f T$. Since in the active case
$\gamma$ depends also on $\tau_R$, we plot $\gamma(f,T)$ for four
different values of the dimensionless observation time
${\cal T}=T/\tau_R$.  We observe that at small $f T$, $\gamma(f,T)$
differs from its BM counterpart and attains a non-universal
$\tau_R$-dependent value when $f T \to 0$. In this region,
$\gamma(f,T)$ is smaller than $\gamma^{(BM)}(f,T)$ and the difference
becomes more pronounced the smaller ${\cal T}$ is. After a minimum
observed at some intermediate value of $f T$, which is only weakly
dependent on ${\cal T}$ (while the minimum itself gets progressively deeper the larger ${\cal T}$ is), we observe a change in the behaviour at large
$fT$ such that $\gamma(f,T)$ becomes larger than
$\gamma^{(BM)}(f,T)$. There, as $f T \to \infty$, $\gamma(f,T)$ tends
to a constant $\tau_R$-dependent value which exceeds $\sqrt{5}/2$, the
limiting value reached by $\gamma^{(BM)}(f,T)$~\cite{krapf1}. Most
strikingly, when ${\cal T}$ increases $\gamma(f,T) \to \sqrt{2}$, away
from the standard BM result.  Overall, Fig.~\ref{fig_gammacfr} reveals
pronounced differences between these two paradigmatic random
processes. Although they both have a similar long-time behaviour if one
looks at the mean-squared displacement only, the coefficient of
variation $\gamma$ is one measurement that allows to distinguish
between the two.

\subsection{Pearson correlation coefficient}

\begin{figure}[htbp]
\centering
\includegraphics[width=100mm]{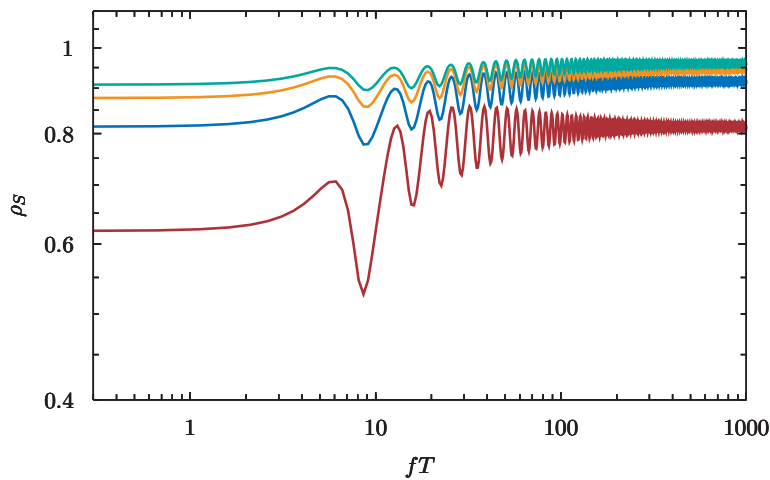}
\caption{Pearson correlation coefficient $\rho_{S}$ as function of
  $fT$ for ${\cal T} = T/\tau_R = 10$ (red), $20$ (blue), $30$
  (orange), and $40$ (green).}
\label{fig_rho}
\end{figure}

The evolution of both components of the ABP position (see
Eqs.~\eqref{02}) originates from the same random process $\theta(t)$
and are thus clearly coupled, which is also obvious from the
requirement that the module of the velocity Eq.~\eqref{v} be fixed.
Clearly, this implies that $S_x(f,T)$ and $S_y(f,T)$ are statistically
correlated. Below, we study the extent of such correlations by
analysing the Pearson coefficient defined in Eq.~\eqref{pearson}. The
behaviour of this coefficient as a function of $f T$ is depicted in
Fig. \ref{fig_rho} for four values of the dimensionless observation
time ${\cal T} = T/\tau_R$.  Physically, such a representation means
that we fix the observation time $T$ and trace the $f$-dependence of
$\rho_S$ for four different values of $\tau_R$ (going from green to
red when increasing $\tau_R$).

We observe first that $S_x(f,T)$ and $S_y(f,T)$ are always positively
correlated since the Pearson coefficient is always positive. Second,
we see that $\rho_S$ is a decreasing function of the persistence time
$\tau_R$. Such a behaviour seems rather counter-intuitive at first
glance since decreasing $\tau_R$ is akin to increasing noise, thus
bringing an ABP closer to standard BM. One could then expect a
decorrelation of the components as $\tau_R$ is reduced. On the other
hand, the correlations between $x$ and $y$ appear precisely because of
the noise. In the extreme limit $\tau_R\to\infty$ the trajectories do
not fluctuate, neither do the STSD and $\rho_s$ thus vanishes. This
gives an argument as to why the Pearson coefficient increases when
$\tau_R$ decreases. Even stronger, our analytical analysis (not shown
here) indicates that $\rho_s$ reaches its maximal value $=1$ in the
limit $T/\tau_R \rightarrow \infty$. The STSD of $x$ and $y$ are then
perfectly correlated.

 \section{Effects of a translational diffusion}
 \label{3}

 In writing the dynamics of an ABP Eq.~(\ref{02}), we have assumed
 that it moves in a solvent imposing a high friction such that it
 reaches a terminal velocity $v$. In principle, the solvent also
 imparts translational noise on the particle. Although the diffusion
 coefficient $D$ originating from such a standard mechanism of
 Brownian motion is typically much smaller than the effective
 diffusion $D_e$ due to self-propulsion \cite{ginot}, the
 presence of a translational diffusion may affect the functional form
 of the spectral density of trajectories. In this Section, we
 investigate how some of the results derived above, especially
 concerning the power spectral density and the coefficient of
 variation, are modified in presence of translational noise.

 We generalise the model of Eqs.~\eqref{02} in a standard way by
 adding to their right-hand-side independent uncorrelated Gaussian
 white-noises so that the trajectories $x_{\rm D}(t)$ and
 $y_{\rm D}(t)$ read
\begin{equation}
\begin{aligned}
\label{a02}
x_{\rm D}(t) & = X_{t} + v \int_{0}^{t}\textrm{d}\tau \, \cos \theta(\tau) \, , \quad x_{\rm D}(0) = 0 \,,  \\
y_{\rm D}(t) & = Y_{t} + v \int_{0}^{t}\textrm{d}\tau \, \sin \theta(\tau) \,, \quad y_{\rm D}(0) = 0 \,,
\end{aligned}
\end{equation}
where $X(t)$ and $Y(t)$ denote independent Brownian motions starting
at the origin at $t=0$, which obey
$\overline{X_{t}} = \overline{Y_t} = 0$ and
$\overline{X_{t} X_{t'}} = \overline{Y_{t} Y_{t'}} = 2 D \, {\rm
  min}\left(t,t'\right)$, and the overbar denotes averaging over these
additional Gaussian noises.

Adding translational noise, the mean-squared displacement for, say, the
$x$-component becomes
\begin{align}
\label{zz}
\overline{\langle x_{\rm D}^{2}(t) \rangle} = 2 D t + v^2 \tau_R^2 \biggl[ \tau - \frac{3}{4} + \frac{2}{3} e^{-\tau} + \frac{1}{12} e^{-4 \tau} \biggr] \,,  \quad \tau = t/\tau_R \,.
\end{align}
The second term in this expression is identical to our previous result
Eq.~\eqref{09}. The translational noise thus appears as a purely
additive contribution. In the limit $t \to \infty$, one finds from
Eq.~\eqref{zz} a standard Brownian behaviour of the form
$\overline{\langle x_{\rm D}^{2}(t) \rangle} \simeq 2 D_{tot} t$ with
$D_{tot} = D_e + D$.  An analogous result for the $y$-component
is obtained by merely replacing the second term in Eq.~\eqref{zz} by the
appropriate expression from Eq.~\eqref{09}.  As a consequence, the
power spectral densities for an infinitely long observation time
$\mu^{\rm D}_x(f)$ and $\mu^{\rm D}_y(f)$ of the processes
$x_{\rm D}(t)$ and $y_{\rm D}(t)$, respectively, after averaging over
the additional white noises, are given by
\begin{align}
\label{tbb}
\mu^{\rm D}_x(f,D) = \mu^{\rm D}_y(f,D) = \frac{2 D_e + 4 D}{f^2} + \frac{2 D_e}{f^2}  \frac{1}{1 + \tau_R^2 f^2} \,.
\end{align}  
Therefore, here again the translational diffusion enters additively
with a contribution corresponding to standard BM with diffusion
constant $D$.

\begin{figure}[htbp]
\centering
\includegraphics[width=100mm]{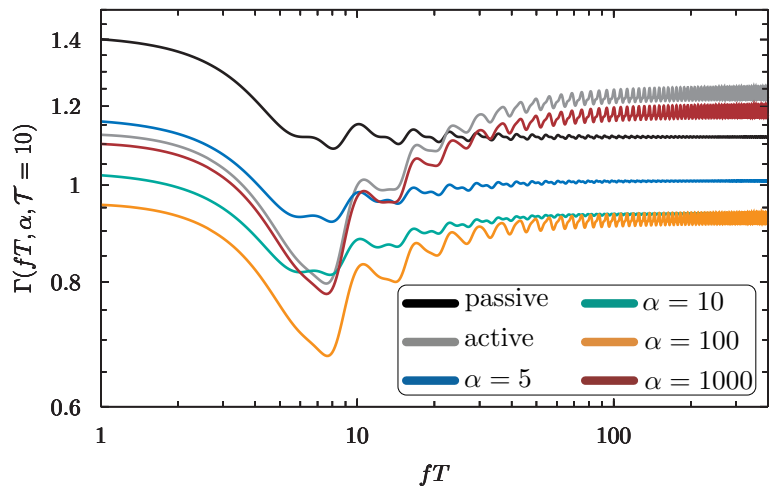}
\caption{The coefficient $\Gamma(fT,\alpha,{\cal T})$ of variation in presence of translation diffusion as function of $f T$ for ${\cal T}=10$ and several values of the parameter
  $\alpha$ defined in Eq.~\eqref{alpha}. The passive case corresponds
  to $\alpha=0$ and the purely active case is reached when
  $\alpha\to\infty$.}
\label{fig_combinedgamma}
\end{figure}

Next, we consider how translational diffusion affects the coefficient
of variation. To this end, let us first note that since the
translational and rotational noises are uncorrelated, the variance
${\rm var} \left(S^{\rm D}_x(f,T,D)\right)$ of the process
$x_{\rm D}(t)$ naturally decomposes into the variance of the
process $x(t)$, which we have studied in the previous Sections, and
the variance of the spectral density of a Brownian motion $X_t$, {\it i.e.}
\begin{align}
{\rm var}\left(S^{\rm D}_x(f,T,D)\right) &= {\rm var}\left(S_x(f,T)\right) + {\rm var}\left(S_X(f,T,D)\right) \nonumber\\
&= \left \langle S^2_x(f,T) \right \rangle - \left \langle S_x(f,T) \right \rangle^2  +  \overline{S^2_X(f,T,D)} - \overline{S_X(f,T,D)}^2 \,.
\end{align}
The second moment of the spectral density of individual trajectories
$x(t)$ has been calculated in Sec.~\ref{2} (see Eq.~\eqref{second} for
its explicit form in the limit $T \to \infty$). The variance
of the spectral density of individual trajectories of a standard
Brownian motion $X_t$ can be found in an explicit form in
Ref.~\cite{krapf1} for arbitrary $T$ and $f$. Capitalising on these
results, we define the coefficient $\Gamma(f,T,D)$ of the process
$x_{\rm D}(t)$:
\begin{align}
\Gamma(f,T, D) = \dfrac{\sqrt{{\rm var}\left(S^{\rm D}_x(f,T,D)\right)}}{\mu^{\rm D}_x(f,T,D)} \,.
\end{align}
Formally, $\Gamma(f,T, D) $ depends on $f$, $T$, $D$, $\tau_{R}$ and
$v$. However, since $\Gamma$ is dimensionless, it must  depend only
on dimensionless combinations of these parameters.  A
straightforward scaling analysis shows that, indeed, $\Gamma(f,T, D) $
is a function of only ${\cal T}=T/\tau_R$, the product $f T$ and the
dimensionless parameter
\begin{equation}
\label{alpha}
\alpha = \frac{v^{2} T}{D} \, .
\end{equation}
The latter interpolates between two limits: $\alpha = 0$ corresponding
to a standard Brownian motion without any self-propulsion, in which
case $ \Gamma(f, T, D)$ reduces to the expression obtained in
Ref.~\cite{krapf1}, and $\alpha = \infty$ -- the case studied in the
previous Sections, {\it i.e.} an active Brownian motion without a
translational diffusion.  The coefficient of variation
$\Gamma(fT,\alpha,{\cal T})$ is plotted in
Fig.~\ref{fig_combinedgamma} for several values of the parameter
$\alpha$ as function of $f T$ including the two limiting cases of
passive BM ($\alpha = 0$) and ABP ($\alpha = \infty$). They exhibit somewhat different trends. For passive BM, $\gamma$
  decreases (with superimposed small oscillations) when $f T$
  increases from the value $\sqrt{2}$ to a smaller value
  $\sqrt{5}/2$. On the contrary, for the purely active motion, the
  coefficient of variation first decreases from $\sqrt{2}$ to a
  minimal value before rising again to the limiting value
  $\sqrt{2}$ in virtue of Eq. \eqref{asymp2}. As a consequence, upon a
  variation of the value of $\alpha$, we observe a change of the trend
  in the behaviour of $\Gamma(fT,\alpha,{\cal T})$ which appears to be
  rather non-trivial due to an interplay between rotational and
  translational diffusion.

More specifically, scaling analysis shows that
  $\Gamma(fT,\alpha,{\cal T})$ exhibits in the limit
  $\alpha \to \infty$ a behaviour of the form
\begin{align}
\label{kl}
\Gamma(fT,\alpha,{\cal T}) = \gamma(f,T) \left(1 - \frac{\beta(f,T)}{\alpha} + O\left(\frac{1}{\alpha^2}\right)\right) \,,
\end{align}
where $\gamma(f,T)$ is the coefficient of variation for an active
Brownian motion and $\beta(f,T)$ is a computable function. Indeed, we
see that for $\alpha = 1000$ (see the magenta curve in
Fig.~\ref{fig_combinedgamma}) $\Gamma(fT,\alpha=1000,{\cal T})$ is
only slightly shifted downwards with respect to the limiting curve
$\Gamma(fT,\alpha=\infty,{\cal T})$. We remark that this case is the
most relevant experimentally since we expect in practice the diffusion
coefficient $D_e$ due to rotational motion and self-propulsion to be
much bigger than the coefficient $D$ of translational diffusion.  For
$\alpha = 100$, we have that $\Gamma(fT,\alpha,{\cal T})$ is still
shifted downwards for any value of the parameter $f T$. However, upon
a further decrease of $\alpha$ the opposite trend establishes -- the
curve $\Gamma(fT,\alpha,{\cal T})$ starts to move upwards (see the
curves corresponding to $\alpha = 10$ and $\alpha = 5$ in
Fig.~\ref{fig_combinedgamma}) and ultimately approaches the black
curve describing the behaviour in the purely Brownian case.
 
 \section{Conclusions}
 \label{4}
 
We analysed the spectral properties of individual
   trajectories of an active Brownian motion, as exemplified by
   the dynamics of a chemically-active Janus colloid confined to move
   on a fluid-fluid interface.
   
   We evaluated the standardly defined power spectral density, which
   is an ensemble-averaged property taken in the limit of an infinite
   observation time $T$.  The resulting expression Eq.\eqref{tb} has a
   more complicated form than the power spectral density of a standard
   Brownian motion (passive case): in addition to a simple power law
   dependence on the frequency $f$, as established for a standard
   Brownian motion, Eq.  \eqref{tb} contains the Lorentzian function
   specific to dynamics in presence of a constant restoring
   force. This causes a departure from the standard Brownian behaviour
   for the intermediate values of $f$, while the large-$f$ and the
   small-$f$ asymptotic behaviours appear to be essentially the same.
   More pronounced differences between the active and the passive case
   are observed for finite observation times -- the more realistic
   situation in experimental and numerical analyses. We have
   demonstrated that for finite $T$ the ensemble-averaged spectral
   density is characterised by a plateau-like region in the short-$f$
   limit, which extends over several orders of magnitude of $f$ and is
   absent for the passive counterpart.
   
   Going beyond the standard approach, we concentrated on fluctuations
   around the ensemble-averaged spectral density and determined the
   coefficient of variation of the distribution of a single-trajectory
   power spectral density. For this property, our analysis revealed
   substantial differences between active and passive Brownian
   motion. Here, the difference is striking not only for finite but
   also infinitely large $T$.  On this basis, we discussed the
   similarities and the distinctive features showing that the spectral
   content of trajectories provides important information and allows
   to distinguish between the two processes.
   
   Finally, we addressed as well the effect of translational diffusion
   on the power spectral density and the coefficient of variation of
   an active Brownian motion. We have shown that in the limit of small
   translational diffusion (compared to the diffusion induced by
   self-propulsion) no singular non-analytic behaviour emerges so that
   the translational diffusion only induced small quantitative
   changes. For larger values of $D$, however, interesting
   non-monotonic behaviours take place.
   
   Our result constitutes a first step in analysing the spectral
   properties of the trajectories of active particles. Going further
   in that direction, it will be important to determine the spectral
   content of other types of active particles, in particular the other
   experimentally relevant  class of run-and-tumble particles.
  

\end{document}